\documentclass[preprint,preprintnumbers, prd, floatfix, superscriptaddress,nofootinbib] {revtex4-1}
\usepackage{epsfig}
\usepackage{subfigure}
\usepackage{dcolumn}% Align table columns on decimal point
\usepackage{bm}% bold math
\usepackage[usenames ,dvipsnames]{xcolor}
\usepackage{slashed}
\usepackage{graphicx,color}
\usepackage[bookmarksnumbered, pdfstartview=FitH,colorlinks,urlcolor=blue, citecolor=blue,linkcolor=blue] {hyperref}
\usepackage{appendix}
\usepackage{amsmath}
\begin{document}

\title{Angular and invariant-mass observables in the four-body Higgs decay
$h\to\ell\bar{\nu}_\ell\bar{\ell}^\prime\nu_{\ell^\prime}$}

\author{Han Zhang}
%\email{Corresponding author: zhanghanzzu@163.com}
\affiliation{School of Physics,
Zhengzhou University, Zhengzhou, Henan 450001, China}

\author{Bai-Cian Ke}
\email{Corresponding author: baiciank@ihep.ac.cn}
\affiliation{School of Physics,
Zhengzhou University, Zhengzhou, Henan 450001, China}

\author{Yao Yu}
\email{Corresponding author: yuyao@cqupt.edu.cn}
\affiliation{Chongqing University of Posts \& Telecommunications, Chongqing, 400065, China}
\affiliation{Department of Physics and Chongqing Key Laboratory for Strongly Coupled Physics, Chongqing University, Chongqing 401331, China}

\author{Yi-Rong Ma}
\email{Corresponding author: mayr@cqupt.edu.cn}
\affiliation{Chongqing University of Posts \& Telecommunications, Chongqing, 400065, China}

\author{Jia-Wei Zhang}
\email{Corresponding author: jwzhang@cqust.edu.cn}
\affiliation{Department of Physics, Chongqing University of Science and Technology, Chongqing, 401331,
 China}

%------------------------------------------------------------------------------
\begin{abstract}
We study the angular distribution of the Higgs boson decay
$h\to\ell\bar{\nu}_\ell\bar{\ell}^\prime\nu_{\ell^\prime}$ with
$\ell\neq\ell^\prime$. Due to the presence of two undetected neutrinos, a
complete angular analysis is not experimentally feasible. To overcome this, we
reorganize the kinematics from the conventional lepton-neutrino pairs into a
charged-lepton pair and a neutrino pair, i.e.~$\ell\bar{\ell}^\prime$ and
$\bar{\nu}_\ell\nu_{\ell^\prime}$. This allows us to express the differential
decay rate in terms of experimentally accessible variables, including the
invariant mass squared of the neutrino pair. Using the effective field theory
framework, we derive this rate and integrate over the neutrino-associated
angles. 
This parametrization provides a clean and measurable angular
distribution, offering a new probe of the $hWW$ coupling and possible
beyond-the-Standard-Model contributions.
\end{abstract}
\maketitle
%------------------------------------------------------------------------------

\section{Introduction}
The discovery of the Higgs boson in 2012 by the ATLAS and CMS
Collaborations~\cite{ATLAS:2012yve, CMS:2012qbp}
has motivated extensive studies of its properties and tests of Standard Model~(SM)
predictions. While no significant discrepancies
from the SM framework have been identified to date, the current experimental
capabilities still leave considerable room for potential
beyond-the-Standard-Model~(BSM) effects. In this context, differential
distributions and kinematic observables are crucial for probing such
potential deviations.

Among various Higgs decay channels, the
$h\to\ell\bar{\nu}_\ell\bar{\ell}^\prime\nu_{\bar{\ell}^\prime}\,(\ell,\ell^\prime=e,\mu,\tau,\,\ell\neq\ell^\prime)$
mode receives contributions only from charged-current amplitudes mediated by
the $W$ boson, without interference from $hZZ$ neutral-current
amplitudes~\cite{Berge:2015jra,LHCHiggsCrossSectionWorkingGroup:2016ypw,Groote:2022pjv,Maina:2020rgd}.
This provides a clean probe of the $hWW$ interaction in the SM and a sensitive
channel for searching for BSM effects~\cite{Flores-Hernandez:2026sac}.
Previous studies by T.~Plehn \textit{et al.}~\cite{Plehn:2001nj, Plehn:1999xi} have
investigated the angular observables in Higgs production processes and the
tensor structure of the Higgs coupling. In contrast, our work focuses on the
decay kinematics.

This four-body Higgs decay is particularly distinctive because of the two neutrinos in the final state.
While the four-momentum of
each individual neutrino cannot be directly accessed in collider experiments
such as ATLAS and CMS, the charged leptons can be identified and measured with
high efficiency and resolution relative to jets. By exploiting four-momentum
conservation, the recoil of these charged leptons
enables the inference of the missing momentum carried by the two neutrinos.
Although the Higgs rest frame may not be fully reconstructable event-by-event
at hadron colliders, our theoretical parametrization is intended to provide
a benchmark for future precision analyses, 
with practical reconstruction
depending on the collider environment and available kinematic
approximation techniques.

In this work, we derive the angular distribution of
$h\to\ell\bar{\nu}_\ell\bar{\ell}^\prime\nu_{\ell^\prime}$.
To better align with measurements, we concentrate on variables that
  are experimentally accessible.
Specifically, the invariant mass
squared of $\bar{\nu}_\ell\nu_{\ell^\prime}$ is taken as
one degree of freedom, and the angular variables involving the
neutrinos are integrated over.

%------------------------------------------------------------------------------

\section{Decay rate formalism}\label{sec:kinematics}
\begin{figure}[t!]
  \centering
  \includegraphics[width=4.0in]{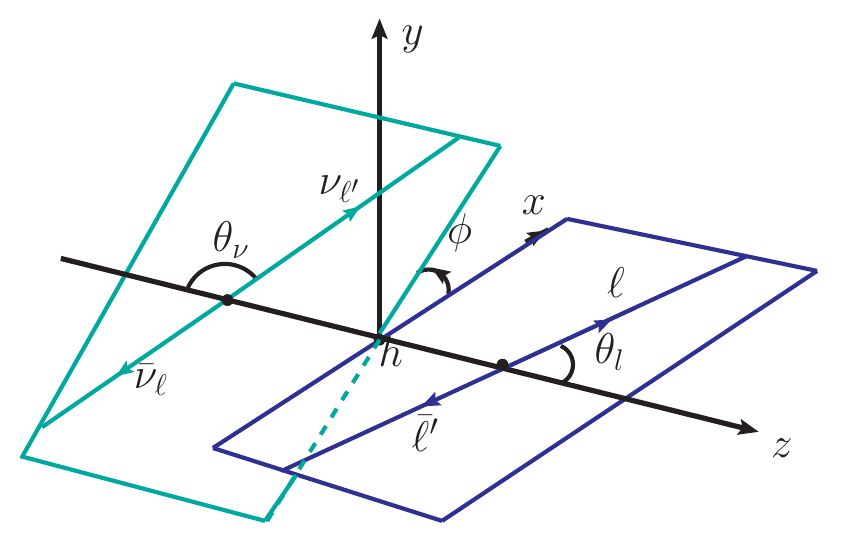}
  \caption{Definition of angles in the decay
    $h\to\ell\bar{\nu}_\ell\bar{\ell}^\prime\nu_{\ell^\prime}$.
    The angle $\theta_L$~($\theta_\nu$) is
    defined as the angle between the three-momentum of
    $\ell$~($\nu_{\ell^\prime}$) in the
    $\ell\bar{\ell}^{\prime}$~($\bar{\nu}_\ell\nu_{\ell^\prime}$) rest frame
    and the flight direction of the
    $\ell\bar{\ell}^{\prime}$~($\bar{\nu}_\ell\nu_{\ell^\prime}$) system in
    the $h$ rest frame; $\phi$ is the angle between the
    $\ell\bar{\ell}^{\prime}$ and the $\bar{\nu}_\ell\nu_{\ell^\prime}$
    planes.}
  \label{fig:angle}
\end{figure}

This section presents the analytical framework required for deriving the
angular distribution of
$h\to\ell\bar{\nu}_\ell\bar{\ell}^\prime\nu_{\ell^\prime}$. Starting from the
charged-current contribution to the three-point function of the Higgs boson
with two fermion currents, we then consider the most general kinematic
description of this four-body decay. The reference systems illustrated in
Fig.~\ref{fig:angle} are adopted for the calculation, and the kinematics is
described in terms of the following five independent variables:
\begin{itemize}
\item $m_{\ell\bar{\ell}^{\prime}}^2$: the invariant mass squared of the
  $\ell\bar{\ell}^{\prime}$ system.
\item $m_{\bar{\nu}_\ell\nu_{\ell^\prime}}^2$: the invariant mass squared of the
  $\bar{\nu}_\ell\nu_{\ell^\prime}$ system.
\item $\theta_L$: the angle between the three-momentum of $\ell$ in the
  $\ell\bar{\ell}^{\prime}$ rest frame and the flight direction of the
  $\ell\bar{\ell}^{\prime}$ system in the $h$ rest frame.
\item $\theta_\nu$: the angle between the three-momentum of $\nu_{\ell^\prime}$
  in the $\bar{\nu}_\ell\nu_{\ell^\prime}$ rest frame and the flight direction
  of the $\bar{\nu}_\ell\nu_{\ell^\prime}$ system in the $h$ rest frame.
\item $\phi$: the angle between the decay planes formed by the
  $\ell\bar{\ell}^{\prime}$ and $\bar{\nu}_\ell\nu_{\ell^\prime}$ momenta
  in the $h$ rest frame with $\phi$
measured from the
  $\ell\bar{\ell}^{\prime}$ plane to the $\bar{\nu}_\ell\nu_{\ell^\prime}$
  plane.
\end{itemize}

%------------------------------------------------------------------------------
\subsection{Amplitude}
Within the Effective Field Theory approach, the charged-current contribution to
the $h\to\ell\bar{\nu}_\ell\bar{\ell}^\prime\nu_{\ell^\prime}$ amplitude can be
written as~\cite{Gonzalez-Alonso:2014eva,Anderson:2013afp, Plehn:2001nj, Plehn:1999xi}
\begin{align}\label{eq:Amplitude}
\mathcal{A}_{c.c.}
\left[
h\to \ell\bar{\nu}_\ell\bar{\ell}^\prime\nu_{\ell^\prime}
\right]
&=
i\,\frac{m_W^2}{2v_F}
\bar{u}_\ell \gamma_\alpha(1-\gamma_5)v_{\bar{\nu}_{\ell}}\bar{u}_{\nu_{\ell^\prime}} \gamma_\beta(1-\gamma_5)v_{\bar{\ell}^\prime}
T^{\alpha\beta}(q_1,q_2),
\\[1ex]
T^{\alpha\beta}(q_1,q_2)
&=
\Bigg[
G_{1}\left(q_1^2,q_2^2\right)\,g^{\alpha\beta}
+ G_{3}\left(q_1^2,q_2^2\right)\,
\frac{q_1\!\cdot\! q_2\, g^{\alpha\beta}
- q_2^\alpha q_1^\beta}{m_W^2}
\nonumber\\
&\qquad
+ G_{4}\left(q_1^2,q_2^2\right)\,
\frac{\epsilon^{\alpha\beta\mu\nu}
q_{2\mu} q_{1\nu}}{m_W^2}
\Bigg],
\end{align}
where $v_F = (\sqrt{2}G_F)^{-1/2}$ with $G_F$ the Fermi constant, $m_W$ is the
$W$ boson mass, and $q_1 \equiv p_\ell + p_{\bar{\nu}_{\ell}}$,
$q_2 \equiv p_{\bar{\ell}^\prime} + p_{\nu_{\ell^\prime}}$.
%------------------------------------------------------------------------------

Assuming the absence of new light states and neglecting contributions from
$D > 6$ operators, the form factors can be decomposed in full generality as
follows:
\begin{align}
  G_{1}\left(q_1^2,q_2^2\right) &= \kappa_{WW}\frac{\left(g^\ell_W\right)^*g^{\ell^\prime}_W}{P_W\left(q_1^2\right)P_W\left(q_2^2\right)}
  +\frac{\left(\epsilon_{W\ell}\right)^*}{m_W^2}\frac{g^{\ell^\prime}_W}{P_W\left(q_2^2\right)}
  +\frac{\epsilon_{W\ell^\prime}}{m_W^2}\frac{\left(g^\ell_W\right)^*}{P_W\left(q_1^2\right)}\,,\nonumber\\
  G_{3}\left(q_1^2,q_2^2\right) &= \epsilon_{WW}\frac{\left(g^\ell_W\right)^*g^{\ell^\prime}_W}{P_W\left(q_1^2\right)P_W\left(q_2^2\right)}\,,\nonumber\\
  G_{4}\left(q_1^2,q_2^2\right) &= \epsilon^{\mathrm{CP}}_{WW}\frac{\left(g^\ell_W\right)^*g^{\ell^\prime}_W}{P_W\left(q_1^2\right)P_W\left(q_2^2\right)}\,,
\end{align}
where $g_W$ are the effective on-shell couplings of the $W$ boson to fermions
and $P_W(q^2)=q^2-m_W^2+im_W\Gamma_W$ is the $W$ propagator.
Following
Refs.~\cite{Anderson:2013afp,Hernandez-Juarez:2024zpk,Hernandez-Juarez:2023dor},
$\kappa_{WW}$, $\epsilon_{WW}$, and $\epsilon_{WW}^{CP}$ are effective dimensionless
couplings that parameterize the possible deviations from the SM $WW$ interaction:
$\kappa_{WW}$ represents the CP-even contribution associated with the SM
left-handed charged-current interaction; $\epsilon_{WW}$ describes additional CP-even
anomalous contributions; $\epsilon_{WW}^{CP}$ denotes the CP-violating coupling.
We assume that $\kappa_{WW}$ is a real coupling, while $\epsilon_{WW}$,
$\epsilon^{\mathrm{CP}}_{WW}$, and $\epsilon_{W\ell^{(\prime)}}$ are generally
allowed to be complex. In the SM at tree level, one has $\kappa_{WW}=1$,
whereas $\epsilon_{WW}$, $\epsilon^{\mathrm{CP}}_{WW}$, and
$\epsilon_{W\ell^{(\prime)}}$ all vanish.

The spinors in
Eq.~(\ref{eq:Amplitude}) are written in pairs of $\ell\bar{\nu}_\ell$ and
$\bar{\ell}^\prime\nu_{\ell^\prime}$. To allow
$m_{\bar{\nu}_\ell \nu_{\ell^\prime}}$ to appear explicitly as a kinematic
variable, the spinors are reorganized into pairs of
$\bar{\nu}_\ell\nu_{\ell^\prime}$ and $\ell\bar{\ell}^\prime$ using the
contraction formula~\cite{Pal:2007dc}
%------------------------------------------------------------------------------
\begin{eqnarray}\label{eq:spinor}
  \bar{u}_\ell \gamma_\alpha(1-\gamma_5)v_{\bar{\nu}_{\ell}}\bar{u}_{\nu_{\ell^\prime}} \gamma_\beta(1-\gamma_5)v_{\bar{\ell}^\prime} &=& \frac{1}{2} \bar{u}_{\nu_{\ell^\prime}}\gamma^\kappa(1-\gamma_5)v_{\bar{\nu}_{\ell}}\bar{u}_\ell \gamma_\alpha \gamma_\kappa\gamma_\beta(1-\gamma_5)v_{\bar{\ell}^\prime}\,.
\end{eqnarray}
The above equation is obtained by applying the Fierz rearrangement to the product of
  two left-handed leptonic currents. Unlike the commonly used chiral Fierz
  identity written in the compact form with contracted Lorentz indices, we
  explicitly rearrange the spinor contractions while keeping track of the
  Lorentz structure. After the Lorentz indices are contracted, the resulting
  expression reduces to the above equation.
The following notations are also introduced for later use:
\begin{align}\label{eq:kw}
k_1 &\equiv p_\ell + p_{\bar{\ell}^\prime}, &
k_2 &\equiv p_\ell - p_{\bar{\ell}^\prime}, \nonumber\\
w_1 &\equiv p_{\nu_{\ell^\prime}} + p_{\bar{\nu}_{\ell}}, &
w_2 &\equiv p_{\nu_{\ell^\prime}} - p_{\bar{\nu}_{\ell}}.
\end{align}

%------------------------------------------------------------------------------
To relate the rest frames of Higgs boson, $\bar{\nu}_\ell\nu_{\ell^\prime}$,
and $\ell\bar{\ell}^\prime$, sets of polarization vectors $\epsilon^\mu(m)$ are
introduced as helicity bases. These polarization vectors satisfy the
orthonormality and completeness properties given by
\begin{eqnarray}
  &&\epsilon^\dagger_{\mu}(m)\epsilon^{\mu}(n)=g_{mn}\,\,\,\,(m,n=t,+,-,0)\,,\nonumber\\
  &&\epsilon^\mu(m)\epsilon^{\dagger\nu}(n)g_{m n}=g^{\mu\nu}\,,
\label{eq:orthonormality}
\end{eqnarray}
where
$g_{m n}={\rm diag}(+, -, -, -)={\rm diag}(g_{tt}, g_{++}, g_{--}, g_{00})$.
In the Higgs boson rest frame, the polarization vectors for the
$\ell\bar{\ell}^\prime$ and $\bar{\nu}_\ell\nu_{\ell^\prime}$ systems can be
written as
\begin{eqnarray}\label{eq:pvector1}
  \epsilon^{\mu}_{\ell\bar{\ell}^\prime,h}(t)
  &=& \frac{1}{\sqrt{k_1^2}}\left(k^0_1,0,0,|\vec{k}_1|\right)\,,\nonumber\\
  \epsilon^{\mu}_{\ell\bar{\ell}^\prime,h}(\pm)
  &=&\mp\frac{1}{\sqrt{2}}\left(0,1,\pm i,0\right)\,,\nonumber\\
  \epsilon^{\mu}_{\ell\bar{\ell}^\prime,h}(0)
  &=& \frac{1}{\sqrt{k_1^2}}\left(|\vec{k}_1|,0,0,k^0_1\right)\,,
\end{eqnarray}
and
\begin{eqnarray}
  \epsilon^{\mu}_{\bar{\nu}_\ell\nu_{\ell^\prime},h}(t)
  &=& \frac{1}{\sqrt{w_1^2}}\left(w^0_1,0,0,-|\vec{k}_1|\right)\,,\nonumber\\
  \epsilon^{\mu}_{\bar{\nu}_\ell\nu_{\ell^\prime},h}(\pm)
  &=& \mp\frac{1}{\sqrt{2}}\left(0,1,\mp i,0\right)\,,\nonumber\\
  \epsilon^{\mu}_{\bar{\nu}_\ell\nu_{\ell^\prime},h}(0)
  &=& \frac{1}{\sqrt{w_1^2}}\left(|\vec{k}_1|,0,0,-w^0_1\right)\,,
\end{eqnarray}
Similarly, the polarization vectors for the
$\ell\bar{\ell}^\prime$ and $\bar{\nu}_\ell\nu_{\ell^\prime}$ systems in their
respective rest frames read
%------------------------------------------------------------------------------
\begin{eqnarray}
  \epsilon^{\mu}_{\ell\bar{\ell}^\prime,\ell\bar{\ell}^\prime}(t)   &=& (1,0,0,0)\,,\nonumber\\
  \epsilon^{\mu}_{\ell\bar{\ell}^\prime,\ell\bar{\ell}^\prime}(\pm)
  &=& \mp\frac{1}{\sqrt{2}} (0,1,\pm i,0)\,,\nonumber\\
  \epsilon^{\mu}_{\ell\bar{\ell}^\prime,\ell\bar{\ell}^\prime}(0)   &=& (0,0,0,1)\,.
\end{eqnarray}
and
\begin{eqnarray}\label{eq:pvector2}
  \epsilon^{\mu}_{\bar{\nu}_\ell\nu_{\ell^\prime},\bar{\nu}_\ell\nu_{\ell^\prime}}(t)
  &=& (1,0,0,0)\,,\nonumber\\
  \epsilon^{\mu}_{\bar{\nu}_\ell\nu_{\ell^\prime},\bar{\nu}_\ell\nu_{\ell^\prime}}(\pm)
  &=&\mp\frac{1}{\sqrt{2}} (0,1,\mp i,0)\,,\nonumber\\
  \epsilon^{\mu}_{\bar{\nu}_\ell\nu_{\ell^\prime},\bar{\nu}_\ell\nu_{\ell^\prime}}(0)
  &=& (0,0,0,-1)\,.
\end{eqnarray}
%------------------------------------------------------------------------------
In addition, one will need the four-momenta $w_2$ in the
$\bar{\nu}_\ell\nu_{\ell^\prime}$ rest frame and $k_2$ in the
$\ell\bar{\ell}^\prime$ rest frame, which are given by
\begin{eqnarray}\label{eq:w2k2}
w_2^{\mu} &=& \left(0, 2|\vec{p}_\nu|\sin\theta_\nu\cos\phi, 2|\vec{p}_\nu|\sin\theta_\nu\sin\phi, -2|\vec{p}_\nu|\cos\theta_\nu\right)\,,\nonumber\\
k_2^{\mu} &=& \left(E_\ell-E_{\bar{\ell}^\prime}, 2\left|\vec{p}_\ell\right|\sin\theta_L, 0, 2\left|\vec{p}_\ell\right|\cos\theta_L\right)\,.
\end{eqnarray}

%------------------------------------------------------------------------------
With the aid of the polarization vectors in
Eqs.~\ref{eq:pvector1}--\ref{eq:pvector2}, together with the kinematic
relations in Eqs.~\ref{eq:kw} and~\ref{eq:w2k2}, the following analytical
expressions for the Lorentz-invariant products involving the four-vectors
$k_{1,2}$ and $w_{1,2}$ are obtained
\begin{align}\label{eq:products}
  k_1 \cdot w_1 &= \frac{m_h^2 - k_1^2 - w_1^2}{2}, \nonumber \\
  k_1 \cdot w_2 &= m_h |\vec{k}_1| \cos\theta_\nu, \nonumber \\
  k_2 \cdot w_1 &= \left(m_{\ell}^2 - m_{\ell^\prime}^2\right) \frac{m_h^2 - k_1^2 - w_1^2}{2k_1^2}
  + 2\left|\vec{p}_\ell\right|\cos\theta_L \frac{m_h |\vec{k}_1|}{\sqrt{k_1^2}}, \nonumber \\
  k_2 \cdot w_2 &= \left(m_{\ell}^2 - m_{\ell^\prime}^2\right) \cos\theta_\nu \frac{m_h |\vec{k}_1|}{k_1^2}
  + \cos\theta_\nu \left|\vec{p}_\ell\right| \cos\theta_L \frac{m_h^2 - k_1^2 - w_1^2}{\sqrt{k_1^2}} \nonumber \\
  &\quad - 2 \sqrt{w_1^2} \sin\theta_\nu \left|\vec{p}_\ell\right| \sin\theta_L \cos\phi, \nonumber \\
  k_2 \cdot k_2 &= \frac{\left(m_{\ell}^2 - m_{\ell^\prime}^2\right)^2}{k_1^2} - 4 |\vec{p}_{\ell}|^2, \nonumber \\
  w_2 \cdot w_2 &= -w_1^2, \nonumber \\
  \epsilon_{\mu\nu\alpha\beta} k_1^{\mu} k_2^{\nu} w_1^{\alpha} w_2^{\beta}
  &= -4 m_h |\vec{k}_1| |\vec{p}_\nu| |\vec{p}_{\ell}|
  \sin\theta_{\ell} \sin\theta_\nu \sin\phi.
\end{align}
%------------------------------------------------------------------------------
Here, $m_h$ is the mass of the Higgs boson; $\vec{k}_1$, $\vec{p}_\ell$, and
$\vec{p}_\nu$ denote the three-momenta of $k_1$, $p_\ell$, and
$p_{\nu_{\ell^\prime}}$ in the Higgs, $\ell\bar{\ell}^\prime$, and
$\bar{\nu}_\ell\nu_{\ell^\prime}$ rest frames, respectively. Substituting
Eqs.~(\ref{eq:spinor}) and~(\ref{eq:products}) into Eq.~(\ref{eq:Amplitude}), the
amplitude squared ultimately becomes
\begin{eqnarray}\label{eq:amplitude}
&&|{\cal A}(h\to \ell(s)\bar{\nu}_\ell\bar{\ell}^\prime(s^\prime)\nu_{\bar{\ell}^\prime})|^2\nonumber\\
&&=\frac{m_w^4}{4v_F^2}\left[G^\ast_{1}G_{1}F_{11}+(G^\ast_{1}G_{3}+G_{1}G^\ast_{3})F^{+}_{13}+i(G^\ast_{1}G_{3}-G_{1}G^\ast_{3})F^{-}_{13}\right.\nonumber\\
&&+\left.(G^\ast_{1}G_{4}+G_{1}G^\ast_{4})F^{+}_{14}+i(G^\ast_{1}G_{4}-G_{1}G^\ast_{4})F^{-}_{14} \right]\,.
\end{eqnarray}
We omit the terms containing $G_3G_3^{*}$ and $G_4G_4^{*}$, whose contributions
are doubly suppressed because $|G_3|$ and $|G_4|$ are much smaller than
$|G_1|$. The functions $F_{11}$, $F^\pm_{13}$, and $F^\pm_{14}$ are given by
%------------------------------------------------------------------------------
\begin{eqnarray}
  F_{11} &=& 256 p_\ell\cdot p_{\nu_{\ell^\prime}} p_{\bar{\ell}^\prime}\cdot p_{\bar{\nu}_{\ell}}\,,\nonumber\\
  F^+_{13} &=& \frac{128}{m_W^2}\left[2m^2_\ell p_{\bar{\ell}^\prime}\cdot p_{\bar{\nu}_{\ell}}p_{\bar{\ell}^\prime}\cdot p_{\nu_{\ell^\prime}}+2m^2_{\ell^\prime}p_{\ell}\cdot p_{\bar{\nu}_{\ell}}p_{\ell}\cdot p_{\nu_{\ell^\prime}}+2p_{\ell}\cdot p_{\nu_{\ell^\prime}}(p_{\bar{\ell}^\prime}\cdot p_{\bar{\nu}_{\ell}})^{2}\right.\nonumber\\
  &&+2(p_{\ell}\cdot p_{\nu_{\ell^\prime}})^2p_{\bar{\ell}^\prime}\cdot p_{\bar{\nu}_{\ell}}+2p_{\ell}\cdot p_{\bar{\nu}_{\ell}}p_{\ell}\cdot p_{\nu_{\ell^\prime}}p_{\bar{\ell}^\prime}\cdot p_{\nu_{\ell^\prime}}
  +2p_{\ell}\cdot p_{\bar{\nu}_{\ell}}p_{\bar{\ell}^\prime}\cdot p_{\bar{\nu}_{\ell}}p_{\bar{\ell}^\prime}\cdot p_{\nu_{\ell^\prime}}\nonumber\\
  &&\left.-2p_{\ell}\cdot p_{\bar{\ell}^\prime} p_{\ell}\cdot p_{\nu_{\ell^\prime}}p_{\bar{\nu}_{\ell}}\cdot p_{\nu_{\ell^\prime}}-2p_{\ell}\cdot p_{\bar{\ell}^\prime} p_{\bar{\ell}^\prime}\cdot p_{\bar{\nu}_{\ell}}p_{\bar{\nu}_{\ell}}\cdot p_{\nu_{\ell^\prime}}-m^2_\ell m^2_{\ell^\prime} p_{\bar{\nu}_{\ell}}\cdot p_{\nu_{\ell^\prime}}\right]\,,\nonumber\\
  F^-_{13} &=&\frac{256}{m_W^2}(p_{\ell}\cdot p_{\nu_{\ell^\prime}}-p_{\bar{\ell}^\prime}\cdot p_{\bar{\nu}_{\ell}})\epsilon_{\mu\nu\alpha\beta}p^\mu_{\ell}p^\nu_{\bar{\ell}^\prime}p^\alpha_{\bar{\nu}_{\ell}}p^\beta_{\nu_{\ell^\prime}}\,,\nonumber\\
  F^+_{14} &=&\frac{256}{m_W^2}(p_{\ell}\cdot p_{\nu_{\ell^\prime}}+p_{\bar{\ell}^\prime}\cdot p_{\bar{\nu}_{\ell}})\epsilon_{\mu\nu\alpha\beta}p^\mu_{\ell}p^\nu_{\bar{\ell}^\prime}p^\alpha_{\bar{\nu}_{\ell}}p^\beta_{\nu_{\ell^\prime}}\,,\nonumber\\
  F^-_{14} &=&\frac{256}{m_W^2}\left[m^2_\ell p_{\bar{\ell}^\prime}\cdot p_{\bar{\nu}_{\ell}}p_{\bar{\ell}^\prime}\cdot p_{\nu_{\ell^\prime}}-m^2_{\ell^\prime}p_{\ell}\cdot p_{\bar{\nu}_{\ell}}p_{\ell}\cdot p_{\nu_{\ell^\prime}}+p_{\ell}\cdot p_{\nu_{\ell^\prime}}(p_{\bar{\ell}^\prime}\cdot p_{\bar{\nu}_{\ell}})^{2}\right.\nonumber\\
  &&-(p_{\ell}\cdot p_{\nu_{\ell^\prime}})^2p_{\bar{\ell}^\prime}\cdot p_{\bar{\nu}_{\ell}}-p_{\ell}\cdot p_{\bar{\nu}_{\ell}}p_{\ell}\cdot p_{\nu_{\ell^\prime}}p_{\bar{\ell}^\prime}\cdot p_{\nu_{\ell^\prime}}
  +p_{\ell}\cdot p_{\bar{\nu}_{\ell}}p_{\bar{\ell}^\prime}\cdot p_{\bar{\nu}_{\ell}}p_{\bar{\ell}^\prime}\cdot p_{\nu_{\ell^\prime}}\nonumber\\
  &&\left.+p_{\ell}\cdot p_{\bar{\ell}^\prime} p_{\ell}\cdot p_{\nu_{\ell^\prime}}p_{\bar{\nu}_{\ell}}\cdot p_{\nu_{\ell^\prime}}-p_{\ell}\cdot p_{\bar{\ell}^\prime} p_{\bar{\ell}^\prime}\cdot p_{\bar{\nu}_{\ell}}p_{\bar{\nu}_{\ell}}\cdot p_{\nu_{\ell^\prime}}\right]\,.\nonumber\\
\end{eqnarray}
With Eq.~(\ref{eq:products}), $F_{11}$, $F_{13}^{\pm}$, and $F_{14}^{\pm}$ can
be expressed as functions of $\phi$, $\cos\theta_{\nu}$, $\cos\theta_{\ell}$,
$k_1^2$, and $w_1^2$. Consequently,
$\left|{\cal A}(h\to \ell(s)\bar{\nu}_{\ell}\bar{\ell}^{\prime}(s^{\prime})\nu_{\bar{\ell}^{\prime}})\right|^2$
also depends on these same five variables.
%------------------------------------------------------------------------------

\subsection{Kinematics and Decay rate}
Using the distinct reference frames defined in Fig.~\ref{fig:angle}, the
differential decay rate for
$h\to \ell\bar{\nu}_{\ell} \bar{\ell}^{\prime}\nu_{\ell^\prime}$ is given by
\begin{eqnarray}\label{eq:kinematics}
  d\Gamma &=& %\int
  \frac{|\vec{k}_1||\vec{p}_{\ell}||\vec{p}_{\nu}|}{(4\pi)^{6}m_h^2\sqrt{k_1^2}\sqrt{w_1^2}}\left|{\cal A}\left(h\to \ell\bar{\nu}_\ell\bar{\ell}^\prime\nu_{\ell^\prime}\right)\right|^2
d\cos\theta_\nu d\phi d\sqrt{k_1^2}d\sqrt{w_1^2}d\cos\theta_L,
\end{eqnarray}
where $\vec{k}_1$, $\vec{p}_\ell$, and $\vec{p}_\nu$ can be expressed
analytically as
\begin{eqnarray}
|\vec{k}_{1}| &=& \frac{\sqrt{m_h^4+\left(k_1^2\right)^2+\left(w_1^2\right)^2-2m_h^2k_1^2-2m_h^2w_1^2-2k_1^2w_1^2}}{2m_h}\,, \nonumber\\
|\vec{p}_{\ell}| &=& \frac{\sqrt{\left(k_1^2\right)^2+m^4_{\ell}+m^4_{\ell^\prime}-2k_1^2m^2_{\ell}-2k_1^2m^2_{\ell^\prime}-2m^2_{\ell}m^2_{\ell^\prime}}}{2\sqrt{k_1^2}}\,, \nonumber\\
|\vec{p}_{\nu}| &=& \frac{\sqrt{w_1^2}}{2}\,.
\end{eqnarray}
The region of integration is specified by
\begin{eqnarray}
  0\leq&\phi &\leq 2\pi\,,\nonumber\\
  -1\leq&\cos\theta_\nu&\leq 1\,,\nonumber\\
-1\leq&\cos\theta_L&\leq 1\,,\nonumber\\
\left(m_{\ell}+m_{\ell^\prime}\right)^{2}&\leq k_1^2 \leq& \left(m_h-\sqrt{w_1^2}\right)^{2}\,, \nonumber\\
0&\leq w_1^2 \leq& \left(m_h-m_{\ell}-m_{\ell^\prime}\right)^{2}\,,\nonumber\\
&{\rm or}&\nonumber\\
0&\leq w_1^2 \leq& \left(m_h-\sqrt{k_1^2}\right)^{2}\,,\nonumber\\
\left(m_{\ell}+m_{\ell^\prime}\right)^{2}&\leq k_1^2 \leq& m_h^{2}\,.
\end{eqnarray}

%------------------------------------------------------------------------------
Next, we perform the integration over the angle $\phi$, which is associated
with the neutrino plane and is not directly measurable. To facilitate this
integration, we first express the $W$
boson propagators in terms of the kinematic variables adopted in this work.
Specifically, $P_W(q_{1,2}^2)$ are
rewritten as functions of $k_1^2$, $w_1^2$, $\theta_L$, $\theta_\nu$, and
$\phi$:
\begin{eqnarray}\label{eq:PW_kw}
  P_W\left(q_{1,2}^2\right) &=&q_{1,2}^2-\left(m_W-i\frac{\Gamma_W}{2}\right)^2= c_{1,2}+i m_W \Gamma_W+|\vec{k}_{1}|\sqrt{w_1^2}\sin\theta_\nu\sin\theta_L\cos\phi
\end{eqnarray}
with
\begin{eqnarray}
  c_1 &=&
 m_h \frac{|\vec{k}_{1}||\vec{p}_{\ell}|}{\sqrt{k_1^2}}\cos\theta_L+(m_\ell^2-m_{\ell^\prime}^2)\frac{m_h^2-w_1^2}{4k_1^2}+\frac{m_h^2-w_1^2-k_1^2}{4}+\frac{3m_\ell^2+m_{\ell^\prime}^2}{4}+\frac{\Gamma_W^2}{4}-m_W^2\nonumber\\
  &&-\left[\frac{m_h}{2}|\vec{k}_{1}|\frac{k_1^2+m_\ell^2-m_{\ell^\prime}^2}{k_1^2}+|\vec{p}_{\ell}|\cos\theta_L \frac{m_h^2-k_1^2-w_1^2}{2\sqrt{k_1^2}}\right]\cos\theta_\nu\,,\nonumber\\
  c_2 &=&
 -m_h \frac{|\vec{k}_{1}||\vec{p}_{\ell}|}{\sqrt{k_1^2}}\cos\theta_L-(m_\ell^2-m_{\ell^\prime}^2)\frac{m_h^2-w_1^2}{4k_1^2}+\frac{m_h^2-w_1^2-k_1^2}{4}+\frac{m_\ell^2+3m_{\ell^\prime}^2}{4}+\frac{\Gamma_W^2}{4}-m_W^2\nonumber\\
  &&-\left[\frac{m_h}{2}|\vec{k}_{1}|\frac{-k_1^2+m_\ell^2-m_{\ell^\prime}^2}{k_1^2}+|\vec{p}_{\ell}|\cos\theta_L \frac{m_h^2-k_1^2-w_1^2}{2\sqrt{k_1^2}}\right]\cos\theta_\nu\,.\nonumber\\
\end{eqnarray}
%------------------------------------------------------------------------------
Note that $c_{1,2}$ are independent of $\phi$ and can therefore be treated as
constants in the $\phi$ integration. In addition, several $\phi$-independent
quantities are introduced for later use:
\begin{eqnarray}
  r_{1,2}^2 &=& \sqrt{\left[c_{1,2}^2-m_W^2 \Gamma_W^2-|\vec{k}_{1}|^2w_1^2\sin^2\theta_\nu\sin^2\theta_L\right]^2+4c_{1,2}^2m_W^2 \Gamma_W^2}\,,\nonumber\\
  y_{1,2}^2&=&c_{1,2}^2-m_W^2 \Gamma_W^2-|\vec{k}_{1}|^2w_1^2\sin^2\theta_\nu\sin^2\theta_L\,.
\end{eqnarray}

\begin{figure}[htp!]
  \centering
  \includegraphics[width=4.0in]{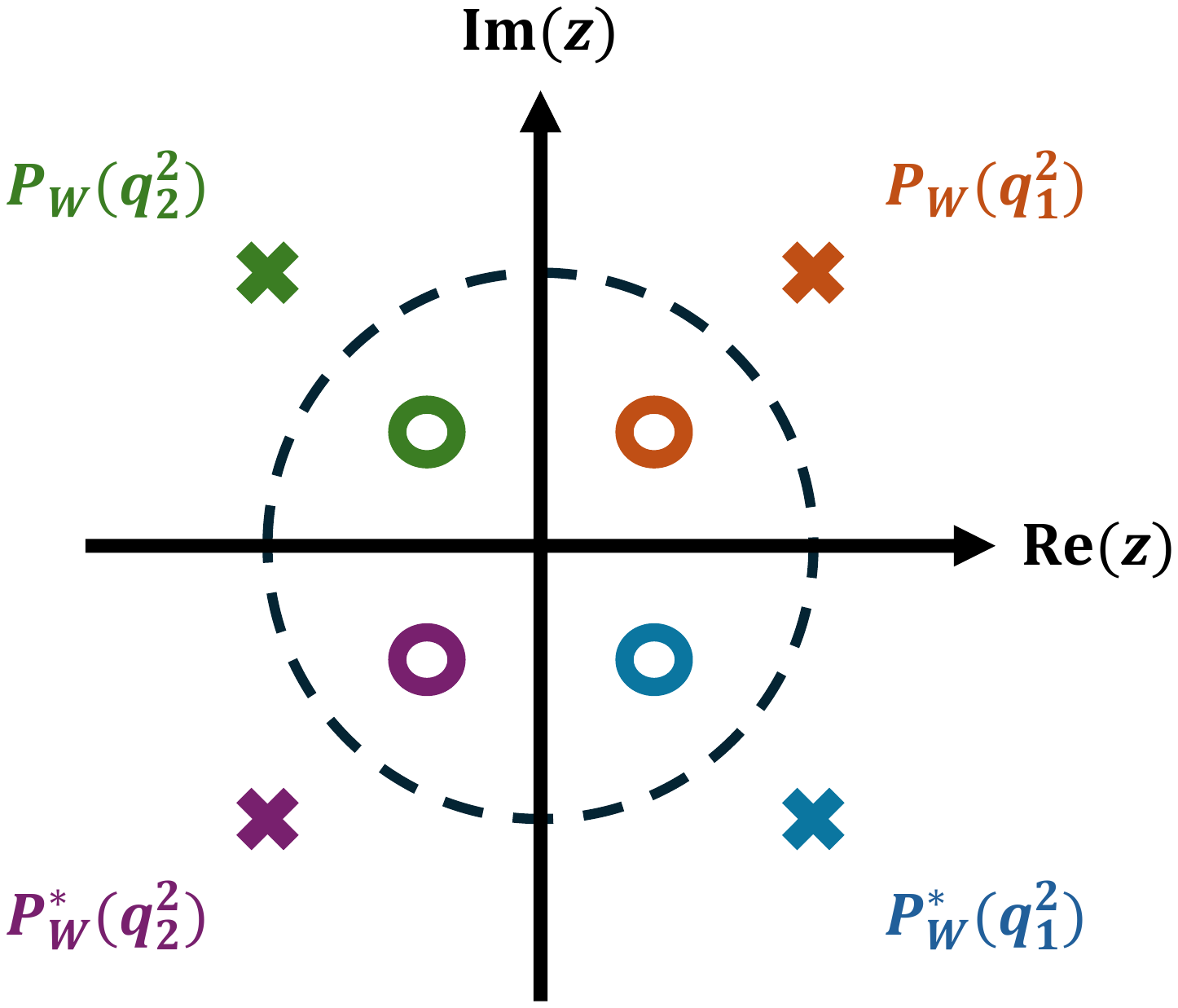}
  \caption{
    Schematic diagram of the pole pairs for each $P_W(q^2)$ in the complex
    plane. Only poles inside the unit circle contribute to the contour
    integration over $\phi$. There is exactly one pole from each pair inside.
    The conjugate partners produce poles reflected across the $\rm{Re}(z)$
    axis.
  }
  \label{fig:root}
\end{figure}

Collecting Eqs.~\ref{eq:Amplitude},~\ref{eq:kinematics} and~\ref{eq:PW_kw},
the differential decay rate integrated over $\phi$ is given by
\begin{eqnarray}\label{eq:residue}
  &&\int_{0}^{2\pi}d\phi\left|{\cal A}\left(h\to \ell\bar{\nu}_\ell\bar{\ell}^\prime\nu_{\ell^\prime}\right)\right|^2\nonumber\\
   &&=\frac{2\pi}{\frac{c_1}{|c_1|}\sqrt{\frac{r_{1}^2+y_1^2}{2}}+i\sqrt{\frac{r_{1}^2-y_1^2}{2}}}\left[P_W(q_{1}^2)\left|{\cal A}\left(h\to \ell\bar{\nu}_\ell\bar{\ell}^\prime\nu_{\ell^\prime}\right)\right|^2\right]\bigg|_{\cos\phi=-\frac{c_{1}+i m_W \Gamma_W}{|\vec{k}_{1}|\sqrt{w_1^2}\sin\theta_\nu\sin\theta_L}}\nonumber\\
   &&+\frac{2\pi}{\frac{c_1}{|c_1|}\sqrt{\frac{r_{1}^2+y_1^2}{2}}-i\sqrt{\frac{r_{1}^2-y_1^2}{2}}}\left[P^*_W\left(q_{1}^2\right)\left|{\cal A}\left(h\to \ell\bar{\nu}_\ell\bar{\ell}^\prime\nu_{\ell^\prime}\right)\right|^2\right]\bigg|_{\cos\phi=-\frac{c_{1}-i m_W \Gamma_W}{|\vec{k}_{1}|\sqrt{w_1^2}\sin\theta_\nu\sin\theta_L}}\nonumber\\
   &&+\frac{2\pi}{\frac{c_2}{|c_2|}\sqrt{\frac{r_{2}^2+y_2^2}{2}}+i\sqrt{\frac{r_{2}^2-y_2^2}{2}}}\left[P_W\left(q_{2}^2\right)\left|{\cal A}\left(h\to \ell\bar{\nu}_\ell\bar{\ell}^\prime\nu_{\ell^\prime}\right)\right|^2\right]\bigg|_{\cos\phi=-\frac{c_{2}+i m_W \Gamma_W}{|\vec{k}_{1}|\sqrt{w_1^2}\sin\theta_\nu\sin\theta_L}}\nonumber\\
   &&+\frac{2\pi}{\frac{c_2}{|c_2|}\sqrt{\frac{r_{2}^2+y_2^2}{2}}-i\sqrt{\frac{r_{2}^2-y_2^2}{2}}}\left[P^*_W\left(q_{2}^2\right)\left|{\cal A}\left(h\to \ell\bar{\nu}_\ell\bar{\ell}^\prime\nu_{\ell^\prime}\right)\right|^2\right]\bigg|_{\cos\phi=-\frac{c_{2}-i m_W \Gamma_W}{|\vec{k}_{1}|\sqrt{w_1^2}\sin\theta_\nu\sin\theta_L}}\,.
\end{eqnarray}
%------------------------------------------------------------------------------
In the above, the residue theorem is employed. The integration over $\phi$ from
$0$ to $2\pi$ is converted into a contour integral along the unit circle in the
complex $z$-plane, with $z = e^{i\phi}$, taken in the counterclockwise
direction. For each $W$ propagator $P_W(q^2)$, there exists a pair of poles,
one lying inside the unit circle and the other outside. A schematic diagram for
the poles of $P_W(q^2)$s is shown in Fig.~\ref{fig:root}. The four propagators
$P_W(q_1^2)$, $P_W^*(q_1^2)$, $P_W(q_2^2)$, and $P_W^*(q_2^2)$ appearing in the
squared amplitude give rise to four poles inside the unit circle, one from each
propagator, which correspond, respectively, to the first through fourth lines
in Eq.~(\ref{eq:residue}).

%------------------------------------------------------------------------------

%------------------------------------------------------------------------------
To evaluate the contributions from the poles, the explicit form of $\sin\phi$
at each pole is required, as it appears in the squared amplitude. At the pole
for $P_W(q_1^2)$, i.e., the first line in Eq.~(\ref{eq:residue}), one has
\begin{equation}
z = \frac{1}{|\vec{k}_{1}| \sqrt{w_1^2}\sin\theta_\nu \sin\theta_L}
\left[
-\left(c_1 + i m_W \Gamma_W\right)
+ \frac{c_1}{|c_1|} \sqrt{\frac{r_1^2 + y_1^2}{2}}
+ i \sqrt{\frac{r_1^2 - y_1^2}{2}}
\right]\,,
\end{equation}
using the relation $z = e^{i\phi}$, or equivalently $\cos\phi = (z^2+1)/(2z)$.
Consequently,
\begin{eqnarray}
  \sin\phi &=& \frac{z^2 - 1}{2iz}
  = \frac{1}{i|\vec{k}_{1}| \sqrt{w_1^2}\sin\theta_\nu \sin\theta_L}\left(\frac{c_1}{|c_1|}\sqrt{\frac{r_{1}^2+y_1^2}{2}}+i\sqrt{\frac{r_{1}^2-y_1^2}{2}}\right)\,.
\end{eqnarray}
The same procedure can be applied to obtain the expressions for $\sin\phi$ at
the other three poles.

%------------------------------------------------------------------------------
\section{NUMERICAL RESULTS AND DISCUSSIONS}
\begin{figure}[htp!]
  \centering
  \includegraphics[width=2.1in]{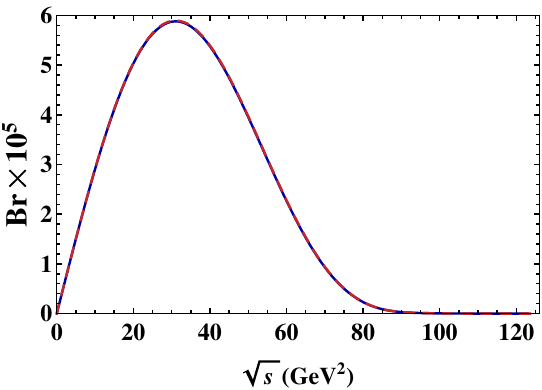}
  \includegraphics[width=2.1in]{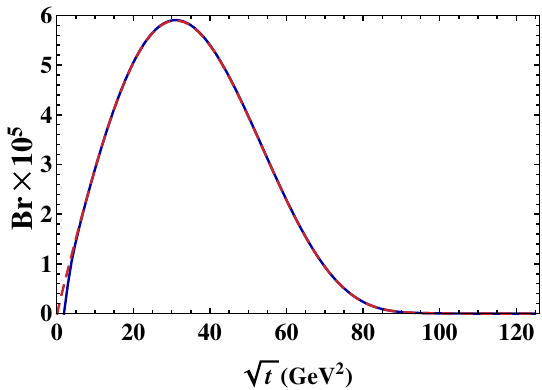}
  \includegraphics[width=2.1in]{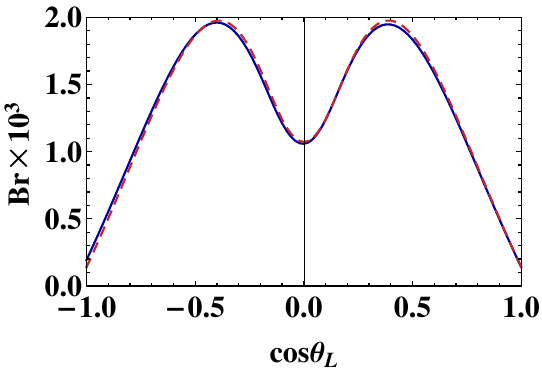}
 \caption{SM tree-level distributions of the branching ratio with respect to
$\sqrt{s}$, $\sqrt{t}$, and $\cos\theta_L$, corresponding to
$\frac{d\mathrm{Br}}{d\sqrt{s}}$, $\frac{d\mathrm{Br}}{d\sqrt{t}}$, and
   $\frac{d\mathrm{Br}}{d\cos\theta_L}$.
The red curves represent $\{\ell, \ell^\prime\}=\{\mu, \bar{e}\}$,
while the blue curves $\{\tau,\bar{e}\}$.
 }
  \label{fig:SM}
\end{figure}

Since neutrinos cannot be detected, we analytically integrate over
$\phi$, and then numerically integrate over $\cos\theta_{\nu}$ to obtain
\begin{eqnarray}
\frac{d\Gamma}{d\sqrt{s}\, d\sqrt{t}\, d\cos\theta_L},
\end{eqnarray}
where $s\equiv w_1^2$ and $t\equiv k_1^2$. Simulated distributions of
the branching ratio as functions of $\sqrt{s}$, $\sqrt{t}$, and
$\cos\theta_L$ are shown in Fig.~\ref{fig:SM}. These results are based
exclusively on the SM tree-level contribution, i.e.~$\kappa_{WW}=1$
with all other couplings set to zero.
The red curves correspond to
$\{\ell, \ell^\prime\}=\{\mu, \bar{e}\}$, and the blue curves to
$\{\tau,\bar{e}\}$. The distributions for $\{\tau,\bar{\mu}\}$ are almost
identical to those for $\{\tau,\bar{e}\}$, and are therefore not shown
explicitly. The remaining charge-conjugated configurations lead to analogous
results and can be related by simple angular transformations.

%------------------------------------------------------------------------------
As for loop-correction and possible BSM effects, we neglect the correction
terms in $G_1$ in the following discussion as they are subdominant compared to
the SM tree-level contribution and exhibit the same kinematic behavior.
To quantify the sensitivity of experimental observables to these effects, we
normalize the corresponding corrections to the SM tree-level contribution.
The normalized correction $\delta_{\mathrm{Br}}$ is defined as
\begin{eqnarray}\label{eq:deltaBr}
\delta_{\mathrm{Br}} &=& \frac{\frac{d\Gamma_{\mathrm{loop, BSM}}}{dY}-\frac{d\Gamma_{\mathrm{SM}}}{dY}}{\frac{d\Gamma_{\mathrm{SM}}}{dY}}\,,
\end{eqnarray}
where $Y=(\sqrt{s},\,\sqrt{t},\,\cos\theta_L)$, $\Gamma_{\mathrm{SM}}$
is the SM decay rate, and
$\Gamma_{\mathrm{loop, BSM}}$ includes
loop-corrections and possible BSM effects.

\begin{figure}[htp!]
  \centering
  \includegraphics[width=2.1in]{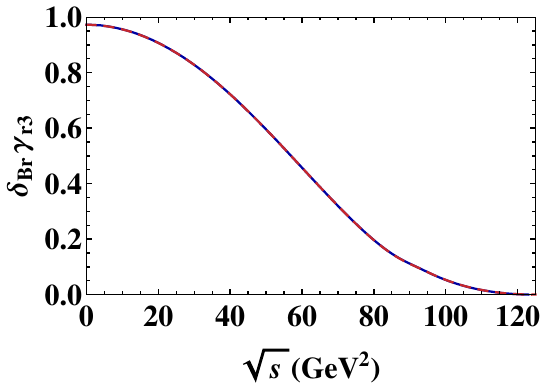}
  \includegraphics[width=2.1in]{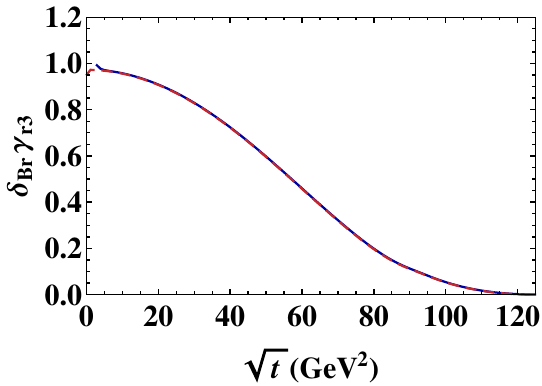}
  \includegraphics[width=2.1in]{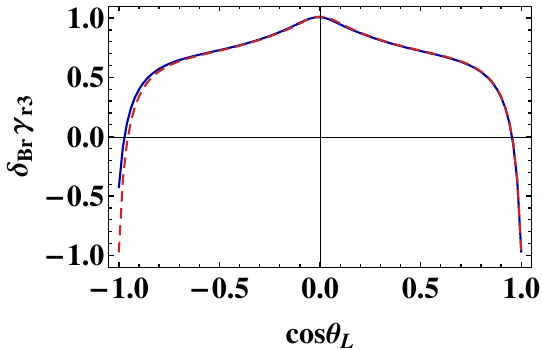}
  \caption{The distributions of $\gamma_{r3}\delta_{\mathrm{Br}}$ as a function
    of $\sqrt{s}$, $\sqrt{t}$, and $\cos\theta_L$. The red curves correspond to
    $\{\ell, \ell^\prime\}=\{\mu, \bar{e}\}$ while the blue curves $\{\tau,\bar{e}\}$.
  }
  \label{fig:3r}
\end{figure}

\begin{figure}[htp!]
  \centering
  \includegraphics[width=2.1in]{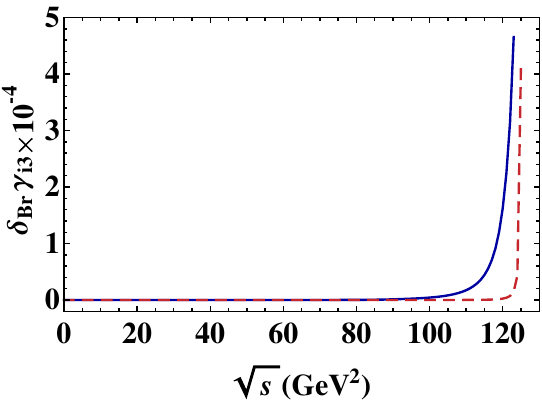}
  \includegraphics[width=2.1in]{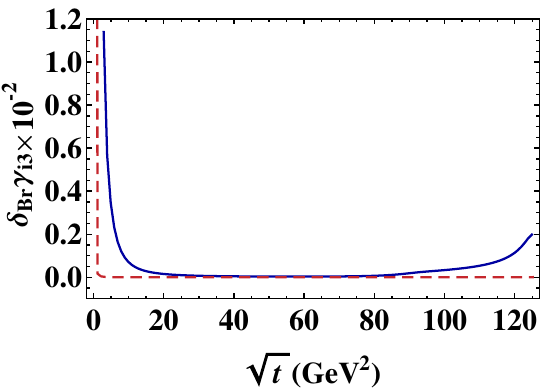}
  \includegraphics[width=2.1in]{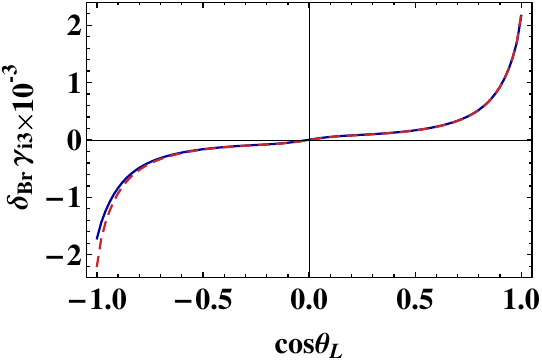}
  \caption{The distributions of $\gamma_{i3}\delta_{\mathrm{Br}}$ as a function
    of $\sqrt{s}$, $\sqrt{t}$, and $\cos\theta_L$. The red curves correspond to
    $\{\ell, \ell^\prime\}=\{\mu, \bar{e}\}$ while the blue curves $\{\tau,\bar{e}\}$.
  }
  \label{fig:3i}
\end{figure}

\begin{figure}[htp!]
  \centering
  \includegraphics[width=2.1in]{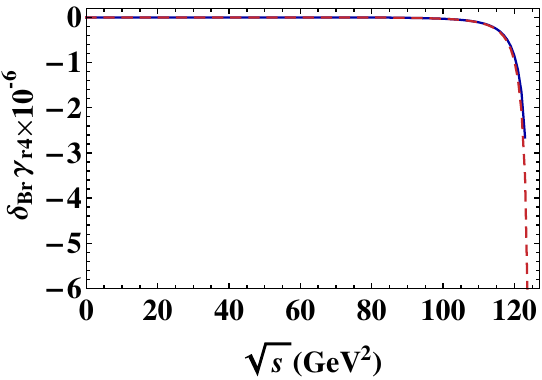}
  \includegraphics[width=2.1in]{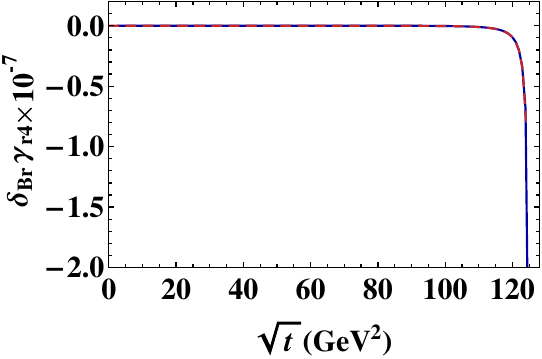}
  \includegraphics[width=2.1in]{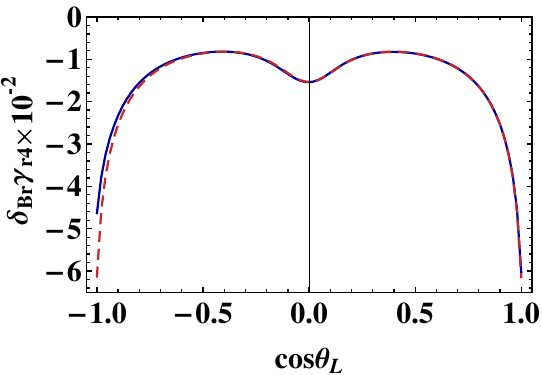}
  \caption{The distributions of $\gamma_{r4}\delta_{\mathrm{Br}}$ as a function
    of $\sqrt{s}$, $\sqrt{t}$, and $\cos\theta_L$. The red curves correspond to
    $\{\ell, \ell^\prime\}=\{\mu, \bar{e}\}$ while the blue curves $\{\tau,\bar{e}\}$.
  }
  \label{fig:4r}
\end{figure}

\begin{figure}[htp!]
  \centering
  \includegraphics[width=2.1in]{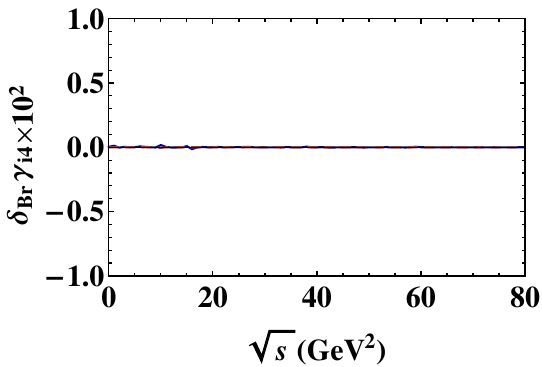}
  \includegraphics[width=2.1in]{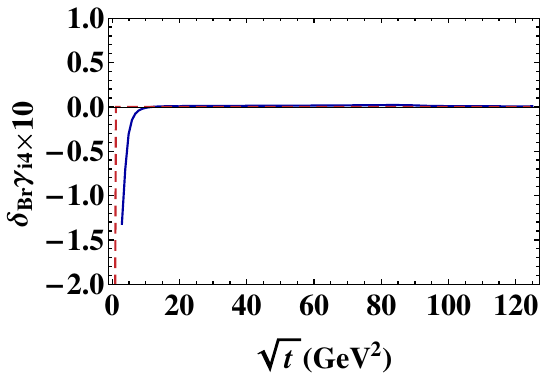}
  \includegraphics[width=2.1in]{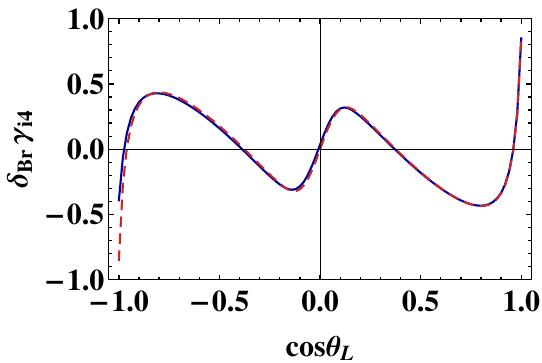}
  \caption{The distributions of $\gamma_{i4}\delta_{\mathrm{Br}}$ as a function
    of $\sqrt{s}$, $\sqrt{t}$, and $\cos\theta_L$. The red curves correspond to
    $\{\ell, \ell^\prime\}=\{\mu, \bar{e}\}$ while the blue curves $\{\tau,\bar{e}\}$.
  }
  \label{fig:4i}
\end{figure}

In general, both the real~(Re) and the imaginary~(Im) parts of $\epsilon_{WW}$ and
$\epsilon^{\mathrm{CP}}_{WW}$ can be non-zero.
We therefore consider four distinct cases: non-zero $\mathrm{Re}[\epsilon_{WW}]$,
non-zero 
$\mathrm{Im}[\epsilon_{WW}]$, non-zero $\mathrm{Re}[\epsilon^{\mathrm{CP}}_{WW}]$,
and non-zero $\mathrm{Im}[\epsilon^{\mathrm{CP}}_{WW}]$, denoted as
$\mathrm{r3}$, $\mathrm{i3}$, $\mathrm{r4}$, and $\mathrm{i4}$, respectively.
Their contributions to $\Gamma_{\mathrm{BSM}}$ arise from the terms involving
$(F_{11},F_{13}^{+})$, $(F_{11},F_{13}^{-})$, $(F_{11},F_{14}^{+})$, and
$(F_{11},F_{14}^{-})$ in Eq.~(\ref{eq:amplitude}), respectively. In contrast,
$\Gamma_{\mathrm{SM}}$ receives contributions only from $F_{11}$.
Furthermore, we define
\[
\gamma_{r3}=\kappa_{WW}/\mathrm{Re}[\epsilon_{WW}]\,, \,
\gamma_{i3}=\kappa_{WW}/\mathrm{Im}[\epsilon_{WW}]\,, \,
\gamma_{r4}=\kappa_{WW}/\mathrm{Re}[\epsilon^{\mathrm{CP}}_{WW}]\,, \,
\gamma_{i4}=\kappa_{WW}/\mathrm{Im}[\epsilon^{\mathrm{CP}}_{WW}]\,.
\]
The simulated distributions of $\gamma_{r3}\delta_{\mathrm{Br}}$,
$\gamma_{i3}\delta_{\mathrm{Br}}$, $\gamma_{r4}\delta_{\mathrm{Br}}$,
and $\gamma_{i4}\delta_{\mathrm{Br}}$ as functions of
$\sqrt{s}$, $\sqrt{t}$, and $\cos\theta_L$ are shown in Figs.~\ref{fig:3r}--\ref{fig:4i}.

The normalized deviation $\delta_{\mathrm{Br}}$, given in Eq.~\ref{eq:deltaBr},
quantifies the relative shift in the differential decay rate induced by
loop-corrections and possible BSM couplings $\epsilon_{WW}$ and $\epsilon_{WW}^{CP}$
with respect to the SM prediction. A larger absolute value
$|\delta_{\mathrm{Br}}|$ in a given kinematic region indicates higher sensitivity
to the corresponding coupling.
The factors $\gamma_{r3}$, $\gamma_{i3}$, $\gamma_{r4}$, and $\gamma_{i4}$ are
introduced to factor out the unknown magnitude of the BSM couplings, so that the
distributions shown in Figs.~\ref{fig:3r}--\ref{fig:4i} represent the relative shape
and sign of the deviation per unit coupling, rather than an absolute cross-section prediction.
For example, a positive $\gamma_{r3}\,\delta_{\mathrm{Br}}$ in a certain bin
means that a positive real part of $\epsilon_{WW}$ would enhance the rate in that
bin relative to the SM, and vice versa. The sensitivity is therefore directly
read from the magnitude of the plotted deviation, with larger
$|\gamma_i\,\delta_{\mathrm{Br}}|$ corresponding to stronger constraining
power on that particular coupling.\\

In Fig.~\ref{fig:3r}, the distributions of the branching ratio with respect to
$\sqrt{s}$, $\sqrt{t}$, and $\cos\theta_L$
exhibit nearly the same sensitivity to $\mathrm{Re}[\epsilon_{WW}]$. Figure~\ref{fig:3i}
indicates $\sqrt{s}$ is the most sensitive observable for probing
$\mathrm{Im}[\epsilon_{WW}]$ in practice. In Fig.~\ref{fig:4r}, the
$\sqrt{t}$ distribution offers the highest sensitivity to
$\mathrm{Re}[\epsilon^{\mathrm{CP}}_{WW}]$. Similarly, Fig.~\ref{fig:4i}
shows that $\cos\theta_L$ is the most sensitive probe for
$\mathrm{Im}[\epsilon^{\mathrm{CP}}_{WW}]$.


\section{Summary}
The decay $h\to\ell\bar{\nu}_\ell\bar{\ell}^\prime\nu_{\ell^\prime}$ with
$\ell\neq\ell^\prime$ is particularly important because it proceeds
purely via charged-current $W$-mediated amplitudes, free from $hZZ$
interference, thus providing a clean probe of the
$hWW$ coupling and a sensitive channel for BSM effects.
In this work, we have derived the angular distribution of this decay in detail.
The presence of two undetected neutrinos renders a conventional angular analysis
in terms of lepton-neutrino pairs infeasible.
We have therefore reorganized the kinematics into a charged-lepton pair~($\ell\bar{\ell}^\prime$)
and a neutrino pair~($\bar{\nu}_\ell\nu_{\ell^\prime}$).
This allows us to express the differential decay rate in terms of
the invariant masses of the $\ell\bar{\ell}^\prime$  and $\bar{\nu}_\ell\nu_{\ell^\prime}$
systems, the helicity angles of these two systems, and the azimuthal angle
between the two decay planes.

Using the effective field theory framework, we have obtained the squared
amplitude and the full differential decay rate. The angle $\phi$, which is
not directly measurable, has been integrated out analytically.
We have presented the numerical results for the distributions of
the invariant mass of $\bar{\nu}_\ell\nu_{\ell^\prime}$, which is experimentally
accessible via four-momentum conservation.
The kinematic reorganization in this work offers a distinct approach compared to
conventional analyses.
Extracting this kinematic structure from data offers a different perspective for testing the SM.

Finally, as in Refs.~\cite{ATLAS:2025hki,ATLAS:2025okx}, the decay channels
studied here are expected to be measured with much higher precision in future
experiments, potentially providing more detailed observables such as angular
distributions that are essential for our analysis. On the theoretical side,
loop-level corrections have also been investigated, as discussed in
Refs.~\cite{Bredenstein:2006rh, Phan:2023xgi, Goncalves:2025xer, Goncalves:2026njf, Greljo:2015sla, Greljo:2017spw}.
These developments together provide
new opportunities for probing BSM physics and for placing stronger constraints on
the parameter space of possible new physics, in particular on the effective
couplings $\epsilon_{WW}$ and $\epsilon^{\mathrm{CP}}_{WW}$.

\section*{ACKNOWLEDGMENTS}
Han Zhang and Bai-Cian Ke were supported in part by National Natural Science Foundation of China~(NSFC) under Contracts No.~12192263, Joint Large-Scale Scientific Facility Fund of the NSFC and the Chinese Academy of Sciences under Contract No.~U2032104, and the Excellent Youth Foundation of Henan Scientific Commitee under Contract No.~242300421044. Yao Yu was supported in part by NSFC under Contracts No.~11905023, No.~12047564 and No.~12147102, the Natural Science Foundation of Chongqing (CQCSTC) under Contract No.~cstc2020jcyj-msxmX0555, and the Science and Technology Research Program of Chongqing Municipal Education Commission (STRPCMEC) under Contracts No.~KJQN202200605 and No.~KJQN202200621; Yi-Rong Ma was supported in part by STRPCMEC under Contracts  No.~KJQN202200650; Jia-Wei Zhang was supported by NSFC under Contract No.~12275036, CQCSTC under Contract  No.~CSTB2025NSCQ-GPX0945, cstc2021jcyjmsxmX0681.

\end{document}